ORIGINAL ARTICLE

# Personalized Forecasting of Glycemic Control in Type 1 and 2 Diabetes Using Foundational AI and Machine Learning Models

Running title: Forecasting glycemic control

Authors:

Simon Lebech Cichosz[1]; Stine Hangaard[1,2]; Thomas Kronborg[1,2]; Peter Vestergaard[2,3]; Morten Hasselstrøm Jensen[1,4]

Author Affiliations:

[1]Department of Health Science and Technology, Aalborg University, Denmark. [2] Steno Diabetes Center North Denmark, Aalborg University Hospital, Denmark. [3] Department of Endocrinology, Aalborg University Hospital, Denmark. [4] Data Science, Novo Nordisk, Søborg, Denmark.

Corresponding author: Simon Cichosz, simcich@hst.aau.dk, Postal address: Selma Lagerløfs Vej 249, 12-02-048, 9260 Gistrup, Danmark, Phone: (+45) 9940 2020; Fax: (+45) 9815 4008. ORCID: 0000-0002-3484-7571

# Abstract

**Background**: Accurate week-ahead forecasts of continuous glucose monitoring (CGM)–derived metrics could enable proactive diabetes management, but relative performance of modern tabular learning approaches is incompletely defined.

**Methods**: We trained and internally validated four regression models (CatBoost, XGBoost, AutoGluon, tabPFN) to predict six week-ahead CGM metrics (TIR, TITR, TAR, TBR, CV, MAGE, and related quantiles) using 4,622 case-weeks from two cohorts (T1DM n=3,389; T2DM n=1,233). Performance was assessed with mean absolute error (MAE) and mean absolute relative difference (MARD); quantile classification was summarized via confusion-matrix heatmaps.

**Results**: Across T1DM and T2DM, all models produced broadly comparable performance for most targets. For T1DM, MARD for TIR, TITR, TAR and MAGE ranged 8.5–16.5% while TBR showed large MARD (mean ≈48%) despite low MAE. AutoGluon and tabPFN showed lower MAE than XGBoost for several targets (e.g., TITR: $p<0.01$; TAR/TBR: $p<0.05$–0.01). For T2DM MARD ranged 7.8–23.9% and TBR relative error was ≈78%; tabPFN outperformed other models for TIR ($p<0.01$), and AutoGluon/ tabPFN outperformed CatBoost/XGBoost on TAR ($p<0.05$). Inference time per 1,000 cases varied markedly (PFN 699 s; AG 2.7 s; CatBoost 0.04 s, XGBoost 0.04 s).

**Conclusions**: Week-ahead CGM metrics are predictable with reasonable accuracy using modern tabular models, but low-prevalence hypoglycemia remains difficult to predict in relative terms. Advanced AutoML and foundation models yield modest accuracy gains at substantially higher computational cost.

# Introduction

Metrics on glycemic control derived from continuous glucose monitoring (CGM) predict all-cause mortality and development of complications (1–4). Future more, the use of CGM has enabled the development of personalized prediction models capable of forecasting short-term glycemic outcomes (5–7) and complications such as gastroparesis and elevated ketone bodies (8–10). In individuals with type 1 diabetes (T1DM), several studies have demonstrated the feasibility of weekly forecasting of clinically relevant metrics such as excessive hypoglycemia, hyperglycemia, and glycemic variability (11–13). While the majority of forecasting research has focused on type 1 diabetes, there is increasing recognition of the potential benefits of CGM in type 2 diabetes (T2DM) (14). Intermittent and continuous CGM use is becoming more widespread in this population, and availability is expected to grow significantly in the coming decade (15).

Accurate glycemic forecasting enables a shift from reactive to proactive diabetes care, which can support earlier interventions and more personalized treatment adjustments (16,17). In this context, prediction models that can forecast glycemic control over longer-period such as the upcoming week are valuable, both for clinical decision-making and for empowering patients in their self-management. However, the applicability and performance of such models in individuals with type 2 diabetes remain underexplored. Moreover, the investigation of prediction precise weekly glycemic control metrics using a regression approach has not been utilized in either type 1 or type 2 diabetes.

Recent advances in machine learning have introduced new modeling approaches that may offer improved predictive performance (18). One such model is Tabular Prior-data Fitted Network (TabPFN) (19), a transformer-based architecture designed specifically for tabular data. TabPFN has been shown to outperform state-of-the-art machine learning techniques, such as ensemble (20), statistical and XGBoost models, across a variety of prediction tasks. Despite these promising results, it remains unclear whether deep neural models like TabPFN can provide meaningful improvements in forecasting glycemic outcomes.

This study aims to evaluate and compare the performance of traditional models, ensemble machine learning approaches, and emerging foundation model for weekly forecasting of glycemic control in individuals with type 1 and type 2 diabetes using CGM data. By identifying the most accurate and reliable modeling strategies, we seek to showcase the potential of a personalized, data-driven tool for diabetes management.

# Methods

## Data material

We developed, trained, and internally validated prediction models using continuous glucose monitoring (CGM) data from two clinical trials, The DiaMonT trial (21,22) and The Insulin-Only Bionic Pancreas Pivotal Trial (IOBP2) (NCT04200313) [14], to estimate the glycemic control (glycemic metrics) in the subsequent weeks.

The DiaMonT trial was a randomized controlled trial evaluating the effectiveness and safety of telemonitoring versus standard care in individuals with type 2 diabetes (T2D) receiving insulin therapy. A total of 331 participants with type 2 diabetes were enrolled (telemonitoring: n = 166; standard care: n = 165). The intervention group used CGM (Dexcom G6), a connected insulin pen, an activity tracker, and smartphone applications for three months. The cohort had a mean age of 61.3 (SD 10.6) years, 61.6% were male, with a median diabetes duration of 16.0 (IQR 12.0) years, and a mean hemoglobin A1c (HbA1c) of 8.01% (SD 1.32) / 64.0 (SD 14.4) mmol/mol.

The IOBP2 trial was a multicenter, randomized controlled trial evaluating an at-home closed-loop system against the standard of care. The study population included individuals with type 1 diabetes (T1D) aged 6 to 79 years. Participants were assigned to either the intervention group, which used the Dexcom G6 continuous glucose monitoring (CGM) system integrated with the iLet Bionic Pancreas for insulin delivery (n = 219), or the control group receiving standard care and CGM (n = 107). The intervention period lasted up to 13 weeks. Overall, 45% of participants were female, the mean HbA1c was 7.8% (SD 1.2), and 89% were CGM users prior to enrollment.

For model development, we included all participants with eligible CGM data. Each prediction case comprised two consecutive weeks with ≥70% CGM coverage. Features from the first week were used to predict outcomes in the second week. Participants without eligible cases were excluded. Data was split into training (70%) and test (30%) sets at the individual level to ensure that no participant contributed to both datasets. The general approach for the methodology is illustrated in Figure 1.

## Prediction targets

Clinical targets for CGM data interpretations are recommending the assessment of both time-in-range (TIR), time-above-range (TAR), time-below-range (TBR) and glycemic variability (23). Hence, based on this recommendation and newer studies (24,25) we selected prediction targets related to time-in-ranges and metrics describing different types of glycemic variability:

- TIR; percentage of time fraction spent at glucose levels 70-180 mg/dl

- Time-in-tight-range (TITR); (26) Percentage of time within a narrower target range (70-140 mg/dL), reflecting tighter glucose control.
- TAR; percentage of time fraction spent at glucose levels above 180 mg/dl
- TBR; percentage of time fraction spent at glucose levels below 70 mg/dl
- Coefficient of Variation (CV): The standard deviation normalized by the mean, expressed as a percentage.
- Mean Amplitude of Glycemic Excursions (MAGE) (27): MAGE captures the average magnitude of significant glucose swings, both increases and decreases, by focusing on excursions that exceed one standard deviation from the mean. It is a widely used indicator of glycemic variability and the likelihood of large glucose fluctuations.

For each case, prediction targets were calculated for the second week.

### Predictors and feature engineering

Prediction of the glycemic outcomes in week two was based on features calculated from CGM data collected during week one for each participant. To derive these features, we employed the open-source MATLAB toolbox *Quantification of Continuous Glucose Monitoring (QoCGM) (28)*, which enables comprehensive analysis of CGM profiles. This approach captures a broad spectrum of information from the glucose traces with potential predictive value.

The derived features encompassed:

- Basic descriptive statistics
- Time-in-Range (TIR) metrics
- Glycemic risk indicators
- Glycemic variability measures (short-, medium-, and long-term)
- Glycemic control indicators
- Entropy and complexity measures

In total, 68 features were generated from the first-week CGM data (see Table 1). Detailed descriptions of the feature calculations are available in the original publication *(28)*. To ensure adequate data quality, only cases with ≥70% CGM coverage were included. Missing data segments were not imputed, in accordance with prior findings suggesting this approach is preferable (29).

### Model development

We employed a regression framework to predict weekly CGM metrics. To provide a comprehensive evaluation, we compared the performance of three modeling strategies: machine learning and stacked ensemble learning methods, and a deep-learning foundation model. All models were trained, tested, and internally validated using identical datasets and methodologies to ensure a fair comparison. Final performance was assessed on the independent test dataset, without recalibration following training.

To enhance generalizability, model tuning was performed through grid search (30) combined with 5-fold cross-validation (31) for parameter optimization. The following models were selected to represent different methodological paradigms and are briefly described below:

- **CatBoost (CB)** (32)**:**
  CB is a gradient boosting algorithm, designed to deliver high performance with minimal parameter tuning. It is particularly well-suited for handling categorical features efficiently. CatBoost incorporates techniques to reduce overfitting, provides robust default parameters, and offers fast training with strong accuracy, making it a popular choice for regression tasks in applied machine learning. In the implementation of CB the hyperparameter: depth, learning rate, iteration, L2 regularization, and subsampling were optimized.

- **Extreme Gradient Boosting (XGB)** (33)**:**
  A tree-based ensemble learning algorithm that builds boosted decision trees in a sequential manner. XGBoost is widely recognized for its robustness, efficiency, and ability to capture complex nonlinear relationships, often outperforming traditional methods in structured tabular data. In the

- **implementation of XBG the hyperparameter: estimators, depth, learning rate, subsampling, regularization alpha and lambda were optimized.**

- **AutoGluon (AG)** (34)**:**
  An automated machine learning (AutoML) framework that integrates multiple algorithms and ensembles them to optimize predictive performance with minimal manual intervention. AutoGluon streamlines model selection and hyperparameter tuning, providing a strong benchmark for automated predictive modeling. The 'hyperparameter' tuned for this model was the time budget.

- **Tabular Prior-Data Fitted Network (tabPFN / PFN)** (35)**:**
  A transformer-based deep learning foundation model pre-trained on synthetic tabular datasets. tabPFN can approximate Bayesian inference without the need for extensive training on the target dataset, enabling fast adaptation and competitive performance even with limited data. The 'hyperparameter' tuned for this model was similar to the AG model the time budget to finetune the model.

To contextualize the performance of the machine learning models, we compared against a simple baseline approach using Last Observation Carried Forward (LOCF). LOCF represents a naïve forecasting strategy in which the most recent observed value is used as the prediction for the subsequent time point. This method requires no model training, makes minimal assumptions, and reflects the level of predictability that can be achieved solely from temporal persistence in the glycemic measures. By including LOCF as a benchmark, we ensure that the machine learning models are evaluated not only in absolute terms but also relative to a clinically intuitive and computationally trivial alternative. (36)

**Model assessment**

The performance of the regression models was evaluated using multiple complementary metrics. Predictive accuracy was quantified by the mean absolute error (MAE) and the mean absolute relative difference (MARD) (Equations 1–2), which capture both absolute and relative deviations between predicted and observed values. The Pearson correlation coefficient (r) and the coefficient of determination ($R^2$) were further calculated to assess the strength of association and the proportion of variance explained by the models, respectively. To formally compare model performance, we applied the Friedman test across models, followed by pairwise Wilcoxon signed-rank tests with Holm–Bonferroni correction in cases where the omnibus test indicated significant differences. Finally, to assess potential clinical utility, we stratified patients into quantile-based risk groups for each metric and evaluated classification accuracy. These results were visualized using enhanced heatmaps of confusion matrices, highlighting each model's ability to correctly assign patients to relevant categories.

*Equation 1*
$$MAE = \frac{1}{n}\sum_{i=1}^{n}|y_i - \hat{y}_i|$$

*Equation 2*
$$MARD = \frac{1}{n}\sum_{i=1}^{n}\frac{|y_i - \hat{y}_i|}{y_i}$$

All analyses were performed using MATLAB (R2021b), Python (v3), the Scikit-learn package (v0.23.2) for machine learning utilities, the autogluon (v1.2), tabpfn (v2.0.5), catboost (v1.2.7) and the XGBoost package (v2.1.1).

# Results

A total of 4,622 case-weeks (T1DM, n = 3,389; T2DM, n = 1,233) from the IOBP2 and DiaMont cohorts were included in the analysis.

For patients with T1DM, predictive performance was generally comparable across the four models for the six targets, as illustrated in Figure 2. The models were able to predict weekly targets with reasonable mean absolute relative difference (MARD) accuracy ranging from 8.5% to 16.5% for TIR, TITR, TAR, and MAGE, with the exception of TBR, which exhibited a mean difference of 48%. This is in contract to the MAE of TBR which is low, but because time spent in hypoglycemia is low, small absolute error will result in a large relative error.

In statistical comparisons, for TITR, all models (CatBoost [CB], AutoGluon [AG], and TabPFN [PFN]) outperformed XGBoost (XGB) with lower mean absolute residuals (MAR, $p < 0.01$). For TAR, AG and PFN demonstrated lower MAR compared with XGB and CB ($p < 0.05$). For TBR, AG and PFN again showed lower MAR compared with XGB and CB ($p < 0.01$). No statistically significant differences were observed for the other targets.

For participants with T2DM, predictive performance was generally comparable across the four models for the six targets, as illustrated in Figure 2. The models were able to predict weekly targets with MARD accuracy ranging from 7.8% to 23.9% for TIR, TITR, TAR, TBR, and MAGE - with TBR, which exhibited a mean difference of 78%. In statistical comparisons between models, for TIR, PFN outperformed CB, AG, XGB (MAR, $p < 0.01$). For TITR and MAGE, XGB had higher mean absolute residuals (MAR, $p < 0.01$) compared to the other models. For TAR. PFN+AG outperformed CB and XGB (MAR, $p < 0.05$). No statistically significant differences were observed for the other targets.

For all glycemic targets and modeling approaches, the LOCF method exhibited significantly higher MARD values ($p < 0.0001$), with correlations between predicted and observed values ranging from 0 to 0.38. Detailed results are provided in Supplementary Tables S1–S2.

The models' ability to classify each target within the correct quantile is summarized in the heatmap-enhanced confusion matrices in Figure 3. Overall, the results from the regression and classification indicate that all models possess substantial capability to predict glycemic control for the following week, although predictions for time in hypoglycemia remain associated with large errors. While marginal differences favor the more complex AG and PFN models, these improvements are not sufficiently robust to suggest large gains in clinical performance. While the training times of more complex models are substantially longer, this represents a one-time computational cost and is therefore not a major limitation. In contrast, the inference time of PFN is markedly slower compared to the other models. For example, in our experiments, prediction of 1,000 cases required 699 seconds with PFN, 2.7 seconds for AG, whereas CatBoost and XGBoost completed the same task in 0.04 seconds when executed on an NVIDIA T1200 GPU (Laptop).

# Discussion

In this study we developed and internally validated four tabular machine-learning regression models (CatBoost, XGBoost, AutoGluon and TabPFN) to predict week-ahead CGM-derived glycemic control metrics in a combined sample of 4,622 case-weeks from the IOBP2 and DiaMont cohorts. Across both diabetes types, models achieved comparable performance for most targets: time in range (TIR), time in tight range (TITR), time above range (TAR) and mean amplitude of glycemic excursions (MAGE) showed reasonable aggregate accuracy. In contrast, relative errors for time below range (TBR) were substantially larger despite low absolute MAE values, an expected statistical consequence when the true target values are near zero. Although AutoGluon and TabPFN produced modest improvements on several endpoints relative to XGBoost and CatBoost, the magnitude of these gains was small to medium and must be weighed against markedly higher computational cost and slower inference. The LOCF approach demonstrated limited predictive capability across all glycemic targets. Its relatively low correlations with observed values and higher error metrics underscore the inherent challenge of forecasting glycemic outcomes based solely on a naïve approach.

This study is, to our knowledge, the first to explore the potential of forecasting glycemic metrics on a week-to-week basic in a regression framework. The main finding is that aggregated, week-level glycemic metrics are predictable from recent CGM inputs with accuracy that may be clinically useful for monitoring and week-to-week planning. Predictability was consistent across multiple modern modeling approaches, suggesting that the underlying CGM-derived features contain stable signal at the weekly horizon. However, more rare or low-prevalence phenomena such as time spent in hypoglycemia remain challenging to predict with low relative error; here event-level detection metrics are more informative for assessing clinical usefulness.

Our results extend two strands of prior research. First, short-horizon glucose forecasting (minutes to hours) is well established and relies on the temporal structure in CGM traces (37–41); we demonstrate that this predictive signal also supports reliable week-level aggregation forecasts, aligning with studies that have modeled weekly glycemic risk prediction from CGM features (11,12,42). Second, recent reports on AutoML and tabular foundation models indicate that these approaches can rival tuned gradient-boosted trees on a range of tabular tasks (19,34,43). Consistent with those reports, TabPFN and AutoGluon in our work occasionally outperformed XGBoost/CatBoost on selected targets; however, the observed advantages were modest. Taken together, these findings suggest a pragmatic pipeline where AutoML/foundation models are considered for rapid prototyping or small-sample problems, while highly optimized gradient-boosted trees remain attractive for production deployment due to favorable runtime and resource profiles.

Week-ahead forecasts of aggregated glycemic metrics have multiple plausible clinical uses. Such as flagging patients at risk of losing time-in-range, prioritizing coaching or clinical outreach, and supporting shared decision making around therapy adjustments. For hypoglycemia, models should be evaluated on clinically meaningful thresholds and event detection. Future work should prioritize external, prospective validation across heterogeneous cohorts and CGM technologies; incorporation of contextual data streams (e.g. insulin dosing, carbohydrate intake, wearable activity measures); and pragmatic trials to measure patient-centered and clinical outcomes when predictions are delivered as decision support.

**Strengths and limitations**

This study benefits from a large, combined dataset spanning both T1DM and T2DM, which enhances the representativeness of the findings. We performed a head-to-head evaluation of multiple contemporary modeling approaches using a consistent preprocessing and evaluation pipeline and reported a comprehensive set of performance measures including absolute and relative error metrics, quantile classification results, and computational metrics.
At the same time, the study has limitations. Evaluation was restricted to internal validation within the IOBP2 and DiaMonT cohorts so external generalizability to other populations, CGM devices, or real-world care settings remain untested; the low base rate of hypoglycemia inflates relative error measures and leaves uncertainty about model performance for clinically important hypoglycemic events. The available feature set likely omitted contextual predictors with potential additional predictive value (for example, precise meal timing, unlogged insulin changes, physical activity, or acute illness). Finally, statistical improvements in prediction do not ensure clinical benefit - randomized or pragmatic trials are needed to determine whether delivering week-ahead forecasts meaningfully changes behavior or improves outcomes.

**Conclusions**

Predicting week-ahead CGM-derived glycemic metrics is feasible with modern tabular machine-learning methods. Most metrics (TIR, TITR, TAR, CV, MAGE) can be forecast with reasonable accuracy, whereas low-prevalence hypoglycemia remains difficult to predict with low relative error. Advanced AutoML and tabular foundation models offer modest performance gains in some settings but incur greater computational costs. External validation and prospective impact studies are required before these models can be recommended for routine clinical use.

**Conflict of Interest:** The research was funded by i-SENS, Inc (Seoul, South Korea) and SLC's involvement with the company did not influence the design, implementation, or interpretation of the study. SLC have received research funding from i-SENS, Inc (Seoul, South Korea), which manufactures some of the product types discussed in this paper. However, the study was conducted independently, and the authors declare that their involvement with i-SENS, Inc (Seoul, South Korea) did not influence the findings or conclusions of the study. PV is head of research at Steno Diabetes Center North Denmark funded by the Novo Nordisk Foundation

**Funding received:** the study was funded by i-SENS, Inc (Seoul, South Korea).

**Disclaimer:** The source of the data is the *Insulin Only Bionic Pancreas Pivotal Trial (NCT04200313)*, but the analyses, content and conclusions presented herein are solely the responsibility of the authors and have not been reviewed or approved by the Bionic Pancreas Research Group or Beta Bionics.

**Ethics statement:** The presented study is a reanalysis of existing and anonymized data from the IOBP / DiaMonT clinical trials. The original study protocols and informed consent forms were approved by the institutional review board(s). Written informed consent was obtained from each participant prior to enrollment of each study. the Regional Ethical Committee of North Jutland, Denmark (N-20200068); ClinicalTrials.gov number(s): NCT04200313, NCT04981808

# References

1. Lu J, Wang C, Shen Y, Chen L, Zhang L, Cai J, et al. Time in Range in Relation to All-Cause and Cardiovascular Mortality in Patients With Type 2 Diabetes: A Prospective Cohort Study. Diabetes Care [Internet]. 2021 Feb 1 [cited 2025 Aug 18];44(2):549–55. Available from: https://dx.doi.org/10.2337/dc20-1862

2. Lu J, Ma X, Zhou J, Zhang L, Mo Y, Ying L, et al. Association of Time in Range, as Assessed by Continuous Glucose Monitoring, With Diabetic Retinopathy in Type 2 Diabetes. Diabetes Care [Internet]. 2018 Nov 1 [cited 2025 Aug 18];41(11):2370–6. Available from: https://dx.doi.org/10.2337/dc18-1131

3. Beck RW, Bergenstal RM, Riddlesworth TD, Kollman C, Li Z, Brown AS, et al. Validation of Time in Range as an Outcome Measure for Diabetes Clinical Trials. Diabetes Care [Internet]. 2019 Mar 1 [cited 2025 Aug 18];42(3):400–5. Available from: https://dx.doi.org/10.2337/dc18-1444

4. Okuno T, Macwan SA, Norman GJ, Miller DR, Reaven PD, Zhou JJ. Continuous Glucose Monitoring Metrics Predict All-Cause Mortality in Diabetes: A Real-world Long-term Study. Diabetes Care [Internet]. 2025 Aug 6 [cited 2025 Aug 18]; Available from: https://dx.doi.org/10.2337/dc25-0716

5. Cichosz SL, Jensen MH, Hejlesen O. Short-term prediction of future continuous glucose monitoring readings in type 1 diabetes: Development and validation of a neural network regression model. Int J Med Inform [Internet]. 2021 Jul [cited 2021 May 4];151:104472. Available from: https://linkinghub.elsevier.com/retrieve/pii/S1386505621000988

6. Fleischer J, Hansen TK, Cichosz SL. Hypoglycemia event prediction from CGM using ensemble learning. Frontiers in Clinical Diabetes and Healthcare. 2022 Dec 9;3:71.

7. Cichosz SL, Frystyk J, Tarnow L, Fleischer J. Combining information of autonomic odulation and CGM measurements nables prediction and improves etection of spontaneous hypoglycemic vents. J Diabetes Sci Technol [Internet]. 2015 Jan 1 [cited 2020 Sep 3];9(1):132–7. Available from: /pmc/articles/PMC4495539/?report=abstract

8. Cichosz SL, Hejlesen O. Classification of Gastroparesis from Glycemic Variability in Type 1 Diabetes: A Proof-of-Concept Study. J Diabetes Sci Technol [Internet]. 2022 Sep 1 [cited 2023 Oct 20];16(5):1190–5. Available from: https://journals.sagepub.com/doi/full/10.1177/19322968211015206

9. Cichosz SL, Bender C. Development of Machine Learning Models for the Identification of Elevated Ketone Bodies During Hyperglycemia in Patients with Type 1 Diabetes. https://home.liebertpub.com/dia [Internet]. 2024 Mar 8 [cited 2024 Mar 12]; Available from: https://www.liebertpub.com/doi/10.1089/dia.2023.0531

10. Ayers AT, Ho CN, Kerr D, Cichosz SL, Mathioudakis N, Wang M, et al. Artificial Intelligence to Diagnose Complications of Diabetes. J Diabetes Sci Technol [Internet]. 2024 Sep 13 [cited 2024 Nov 15]; Available from: https://vbn.aau.dk/da/publications/artificial-intelligence-to-diagnose-complications-of-diabetes

11. Giammarino F, Senanayake R, Prahalad P, Maahs DM, Scheinker D. A Machine Learning Model for Week-Ahead Hypoglycemia Prediction From Continuous Glucose Monitoring Data. https://doi.org/101177/19322968241236208 [Internet]. 2024 Mar 6 [cited 2024 Jun 5]; Available from: https://journals.sagepub.com/doi/abs/10.1177/19322968241236208

12. Cichosz SL, Jensen MH, Olesen SS. Development and Validation of a Machine Learning Model to Predict Weekly Risk of Hypoglycemia in Patients with Type 1 Diabetes Based on Continuous Glucose Monitoring. https://home.liebertpub.com/dia [Internet]. 2024 Jan 12 [cited 2024 Jan 16]; Available from: https://www.liebertpub.com/doi/10.1089/dia.2023.0532

13. Cichosz SL, Olesen SS, Jensen MH. Explainable Machine-Learning Models to Predict Weekly Risk of Hyperglycemia, Hypoglycemia, and Glycemic Variability in Patients With Type 1 Diabetes Based on Continuous Glucose Monitoring. https://doi.org/101177/19322968241286907 [Internet]. 2024 Oct 8 [cited 2024 Oct 24]; Available from: https://journals.sagepub.com/doi/abs/10.1177/19322968241286907

14. Battelino T, Lalic N, Hussain S, Ceriello A, Klobucar S, Davies SJ, et al. The use of continuous glucose monitoring in people living with obesity, intermediate hyperglycemia or type 2 diabetes. Diabetes Res Clin Pract [Internet]. 2025 May 1 [cited 2025 Aug 25];223:112111. Available from: https://www.sciencedirect.com/science/article/pii/S0168822725001251

15. Ziegler R, Heinemann L, Freckmann G, Schnell O, Hinzmann R, Kulzer B. Intermittent Use of Continuous Glucose Monitoring: Expanding the Clinical Value of CGM. J Diabetes Sci Technol [Internet]. 2021 [cited 2025 Aug 25];15(3):684–94. Available from: https://scholar.google.com/scholar_url?url=https://journals.sagepub.com/doi/pdf/10.1177/1932296820905577&hl=da&sa=T&oi=ucasa&ct=usl&ei=rAmsaIqFCpXUieoPmrax2A8&scisig=AAZF9b86s26sbo24VqxRfdpdQX87

16. Klonoff DC, Bergenstal RM, Cengiz E, Clements MA, Espes D, Espinoza J, et al. CGM Data Analysis 2.0: Functional Data Pattern Recognition and Artificial Intelligence Applications. J Diabetes Sci Technol [Internet]. 2025 Aug 14 [cited 2025 Aug 25]; Available from: https://scholar.google.com/scholar_url?url=https://journals.sagepub.com/doi/pdf/10.1177/19322968251353228&hl=da&sa=T&oi=ucasa&ct=usl&ei=twWsaOb1D6zWieoPic2ZoAU&scisig=AAZF9b-k2fzwPWgS5tca0IBboG7w

17. Bender C, Vestergaard P, Cichosz SL. The History, Evolution and Future of Continuous Glucose Monitoring (CGM). Diabetology 2025, Vol 6, Page 17 [Internet]. 2025 Mar 3 [cited 2025 Aug 25];6(3):17. Available from: https://www.mdpi.com/2673-4540/6/3/17/htm

18. Mehdizavareh H, Khan A, Cichosz SL. Enhancing glucose level prediction of ICU patients through hierarchical modeling of irregular time-series. Comput Struct Biotechnol J [Internet]. 2025 Jan 1 [cited 2025 Aug 4];27:2898–914. Available from: https://www.sciencedirect.com/science/article/pii/S2001037025002545

19. Hollmann N, Müller S, Purucker L, Krishnakumar A, Körfer M, Hoo S Bin, et al. Accurate predictions on small data with a tabular foundation model. Nature 2025 637:8045 [Internet]. 2025 Jan 8 [cited 2025 Feb 14];637(8045):319–26. Available from: https://www.nature.com/articles/s41586-024-08328-6

20. Erickson N, Mueller J, Shirkov A, Zhang H, Larroy P, Li M, et al. AutoGluon-Tabular: Robust and Accurate AutoML for Structured Data. 2020 Mar 13 [cited 2025 Aug 6]; Available from: https://arxiv.org/pdf/2003.06505

21. Hangaard S, Kronborg T, Hejlesen O, Aradóttir TB, Kaas A, Bengtsson H, et al. The Diabetes teleMonitoring of patients in insulin Therapy (DiaMonT) trial: study protocol for a randomized controlled trial. Trials [Internet]. 2022 Dec 1 [cited 2024 Apr 16];23(1):1–9. Available from: https://trialsjournal.biomedcentral.com/articles/10.1186/s13063-022-06921-6

22. Hangaard S, Kronborg T, Cohen SR, Kofoed-Enevoldsen A, Thomsen CHN, Aradóttir TB, et al. Effectiveness and Safety of Telemonitoring Compared with Standard of Care in People with Type 2 Diabetes Treated with Insulin: A National Multicenter Randomized Controlled Trial. 2024 [cited 2025 Apr 7]; Available from: https://papers.ssrn.com/abstract=5032694

23. Battelino T, Danne T, Bergenstal RM, Amiel SA, Beck R, Biester T, et al. Clinical targets for continuous glucose monitoring data interpretation: Recommendations from the international consensus on time in range. Diabetes Care [Internet]. 2019 Aug 1 [cited 2021 Jun 29];42(8):1593–603. Available from: https://pubmed.ncbi.nlm.nih.gov/31177185/

24. Beck RW, Raghinaru D, Calhoun P, Bergenstal RM. A Comparison of Continuous Glucose Monitoring-Measured Time-in-Range 70–180 mg/dL Versus Time-in-Tight-Range 70–140 mg/dL. Diabetes Technol Ther [Internet]. 2024 Mar 1 [cited 2025 Aug 15];26(3):151–5. Available from: /doi/pdf/10.1089/dia.2023.0380?download=true

25. Lebech Cichosz S, Kronborg ; Thomas, Laugesen E, Hangaard S, Fleischer J, Troels ;, et al. From Stability to Variability: Classification of Healthy Individuals, Prediabetes, and Type 2 Diabetes using Glycemic Variability Indices from Continuous Glucose Monitoring Data. https://home.liebertpub.com/dia [Internet]. 2024 Aug 8 [cited 2024 Aug

26]; Available from: https://www.liebertpub.com/doi/10.1089/dia.2024.0226

26. Beck RW, Raghinaru D, Calhoun P, Bergenstal RM. A Comparison of Continuous Glucose Monitoring-Measured Time-in-Range 70–180 mg/dL Versus Time-in-Tight-Range 70–140 mg/dL. Diabetes Technol Ther [Internet]. 2024 Mar 1 [cited 2024 Aug 26];26(3):151–5. Available from: https://www.liebertpub.com/doi/10.1089/dia.2023.0380

27. Service FJ, Molnar GD, Rosevear JW, Ackerman E, Gatewood LC, Taylor WF. Mean Amplitude of Glycemic Excursions, a Measure of Diabetic Instability. Diabetes [Internet]. 1970 Sep 1 [cited 2024 Jan 10];19(9):644–55. Available from: https://dx.doi.org/10.2337/diab.19.9.644

28. Cichosz S, Hangaard S, Kronborg T, Vestergaard P, Jensen MH. From data to insights: a tool for comprehensive Quantification of Continuous Glucose Monitoring (QoCGM). PeerJ [Internet]. 2025 Jun 9 [cited 2025 Aug 18];13:e19501. Available from: https://peerj.com/articles/19501

29. Cichosz SL, Kronborg T, Hangaard S, Vestergaard P, Jensen MH. Assessing the Accuracy of Continuous Glucose Monitoring Metrics: The Role of Missing Data and Imputation Strategies. Diabetes Technol Ther [Internet]. 2025 May 14 [cited 2025 Aug 18];00(00):2025. Available from: /doi/pdf/10.1089/dia.2025.0102?download=true

30. Lerman PM. Fitting Segmented Regression Models by Grid Search. J R Stat Soc Ser C Appl Stat [Internet]. 1980 Mar 1 [cited 2023 Jun 7];29(1):77–84. Available from: https://dx.doi.org/10.2307/2346413

31. Fushiki T. Estimation of prediction error by using K-fold cross-validation. Stat Comput [Internet]. 2011 Apr 1 [cited 2023 Feb 14];21(2):137–46. Available from: https://link.springer.com/article/10.1007/s11222-009-9153-8

32. Prokhorenkova L, Gusev G, Vorobev A, Dorogush AV, Gulin A. CatBoost: unbiased boosting with categorical features. Adv Neural Inf Process Syst [Internet]. 2018 [cited 2025 Aug 25];31. Available from: https://github.com/catboost/catboost

33. Chen T, Guestrin C. XGBoost: A Scalable Tree Boosting System. Proceedings of the ACM SIGKDD International Conference on Knowledge Discovery and Data Mining [Internet]. 2016 Mar 9 [cited 2023 Jun 7];13-17-August-2016:785–94. Available from: https://arxiv.org/abs/1603.02754v3

34. Erickson N, Mueller J, Shirkov A, Zhang H, Larroy P, Li M, et al. AutoGluon-Tabular: Robust and Accurate AutoML for Structured Data. 2020 Mar 13 [cited 2025 Feb 14]; Available from: https://arxiv.org/abs/2003.06505v1

35. Hollmann N, Müller S, Purucker L, Krishnakumar A, Körfer M, Hoo S Bin, et al. Accurate predictions on small data with a tabular foundation model. Nature 2025 637:8045 [Internet]. 2025 Jan 8 [cited 2025 Aug

18];637(8045):319–26. Available from: https://www.nature.com/articles/s41586-024-08328-6

36. Dave D, DeSalvo DJ, Haridas B, McKay S, Shenoy A, Koh CJ, et al. Feature-Based Machine Learning Model for Real-Time Hypoglycemia Prediction: https://doi.org/101177/1932296820922622 [Internet]. 2020 Jun 1 [cited 2021 Aug 6];15(4):842–55. Available from: https://journals.sagepub.com/doi/abs/10.1177/1932296820922622

37. Cichosz SL, Frystyk J, Hejlesen OK, Tarnow L, Fleischer J. A Novel Algorithm for Prediction and Detection of Hypoglycemia Based on Continuous Glucose Monitoring and Heart Rate Variability in Patients With Type 1 Diabetes: https://doi.org/101177/1932296814528838 [Internet]. 2014 Mar 31 [cited 2021 Aug 6];8(4):731–7. Available from: https://journals.sagepub.com/doi/full/10.1177/1932296814528838

38. Cichosz SL, Frystyk J, Tarnow L, Fleischer J. Combining Information of Autonomic Modulation and CGM Measurements Enables Prediction and Improves Detection of Spontaneous Hypoglycemic Events: https://doi.org/101177/1932296814549830 [Internet]. 2014 Sep 12 [cited 2021 Aug 6];9(1):132–7. Available from: https://journals.sagepub.com/doi/full/10.1177/1932296814549830

39. Woldaregay AZ, Årsand E, Botsis T, Albers D, Mamykina L, Hartvigsen G. Data-Driven Blood Glucose Pattern Classification and Anomalies Detection: Machine-Learning Applications in Type 1 Diabetes. J Med Internet Res 2019;21(5):e11030 https://www.jmir.org/2019/5/e11030 [Internet]. 2019 May 1 [cited 2021 Nov 12];21(5):e11030. Available from: https://www.jmir.org/2019/5/e11030

40. Woldaregay AZ, Årsand E, Walderhaug S, Albers D, Mamykina L, Botsis T, et al. Data-driven modeling and prediction of blood glucose dynamics: Machine learning applications in type 1 diabetes. Artif Intell Med [Internet]. 2019 Jul 1 [cited 2025 May 15];98:109–34. Available from: https://pubmed.ncbi.nlm.nih.gov/31383477/

41. Cichosz SL, Jensen MH, Olesen SS. Development and Validation of a Machine Learning Model to Predict Weekly Risk of Hypoglycemia in Patients with Type 1 Diabetes Based on Continuous Glucose Monitoring. https://home.liebertpub.com/dia [Internet]. 2024 Jul 12 [cited 2024 Nov 15];26(7):457–66. Available from: https://www.liebertpub.com/doi/10.1089/dia.2023.0532

42. Shchur O, Turkmen C, Erickson N, Shen H, Shirkov A, Hu T, et al. AutoGluon-TimeSeries: AutoML for Probabilistic Time Series Forecasting. Proc Mach Learn Res [Internet]. 2023 Aug 10 [cited 2025 Feb 14];228. Available from: https://arxiv.org/abs/2308.05566v1

| Metric | Description | All | Day | Night |
|---|---|---|---|---|
| Median | Median glucose value | X | X | X |
| Std | Standard deviation of glucose values | X | X | X |
| CV | Coefficient of variation | X | X | X |
| IQR | Interquartile range | X | X | X |
| Pctile75 | 75th percentile of glucose values | X | X | X |
| Pctile25 | 25th percentile of glucose values | X | X | X |
| TIR | Time in range (70-180 mg/dL) | X | X | X |
| TITR | Time in tight range (70-140 mg/dL) | X | X | X |
| TBR1 | Time below range (54-70 mg/dL) | X | X | X |
| TBR2 | Time below range (<54 mg/dL) | X | X | X |
| TBR | Total time below range (TBR1 + TBR2) | X | X | X |
| TAR1 | Time above range (180-250 mg/dL) | X | X | X |
| TAR2 | Time above range (>250 mg/dL) | X | X | X |
| TAR | Total time above range (TAR1 + TAR2) | X | X | X |
| Hypo_episodes_n | Number of hypoglycemia events (<70 mg/dL) | X | | |
| GRI_Hypo | Glucose Risk Index for hypoglycemia | X | | |
| GRI_Hyper | Glucose Risk Index for hyperglycemia | X | | |
| GRI | Glucose Risk Index | X | | |
| CONGA_1H | Continuous Overall Net Glycemic Action over 1 hour | X | | |
| CONGA_2H | Continuous Overall Net Glycemic Action over 2 hours | X | | |
| CONGA_6H | Continuous Overall Net Glycemic Action over 6 hours | X | | |
| CONGA_24H | Continuous Overall Net Glycemic Action over 24 hours | X | | |
| MAGE | Mean Amplitude of Glycemic Excursion | X | | |
| Mobility | Signal mobility | X | | |
| DTpM | Distance traveled per minute | X | | |
| FGxP | Fasting glucose proxy | X | | |
| GMI | Glucose Management Indicator | X | | |
| LBGI | Low Blood Glucose Index | X | | |
| HBGI | High Blood Glucose Index | X | | |
| MCI | Multiscale Complexity Index | X | | |
| GRADE | Glycemic Risk Assessment Diabetes Equation score | X | | |
| GRADE_hypo | Percentage of GRADE score for hypoglycemia | X | | |
| GRADE_eu | Percentage of GRADE score for euglycemia | X | | |
| GRADE_hyper | Percentage of GRADE score for hyperglycemia | X | | |
| D2d_mean | Day-to-day standard deviation of mean glucose | X | | |
| D2d_TIR | Day-to-day standard deviation of time in range | X | | |

**Table 1 –** The figure displays the CGM features calculated by QoCGM with a short description.

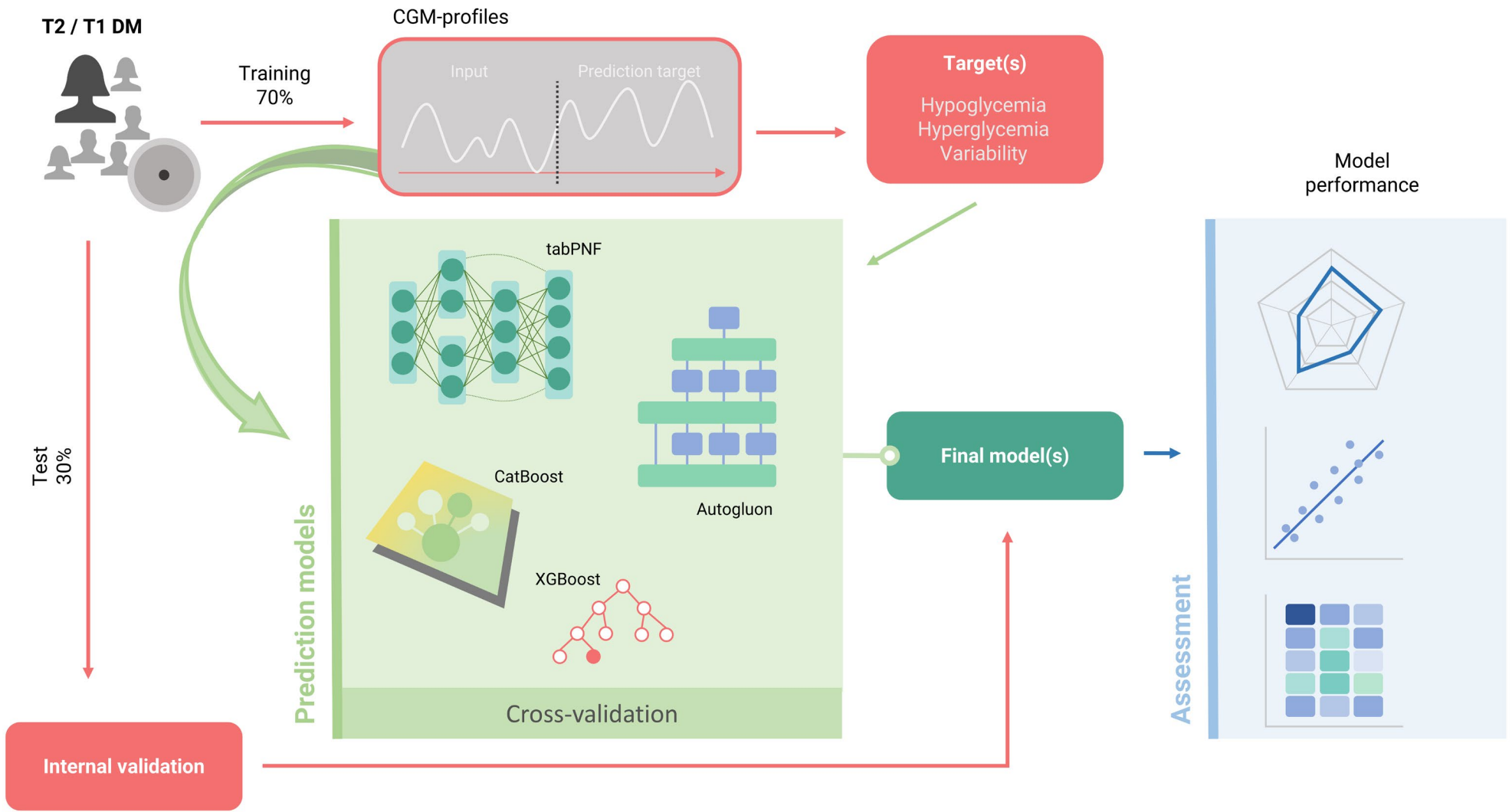

*Figure 1* – Schematic overview of the methodological approach. Continuous glucose monitoring (CGM) profiles from individuals with type 1 and type 2 diabetes were split into training (70%) and test (30%) datasets. Prediction models (tabPFN, CatBoost, AutoGluon, and XGBoost) were trained using cross-validation to forecast next-week glycemic control metrics. The models targeted included time-in-ranges such as hypoglycemia, hyperglycemia, and glucose variability outcomes. Final models were evaluated on the test set through internal validation, with performance assessed using correlation, error metrics, and visualization techniques.

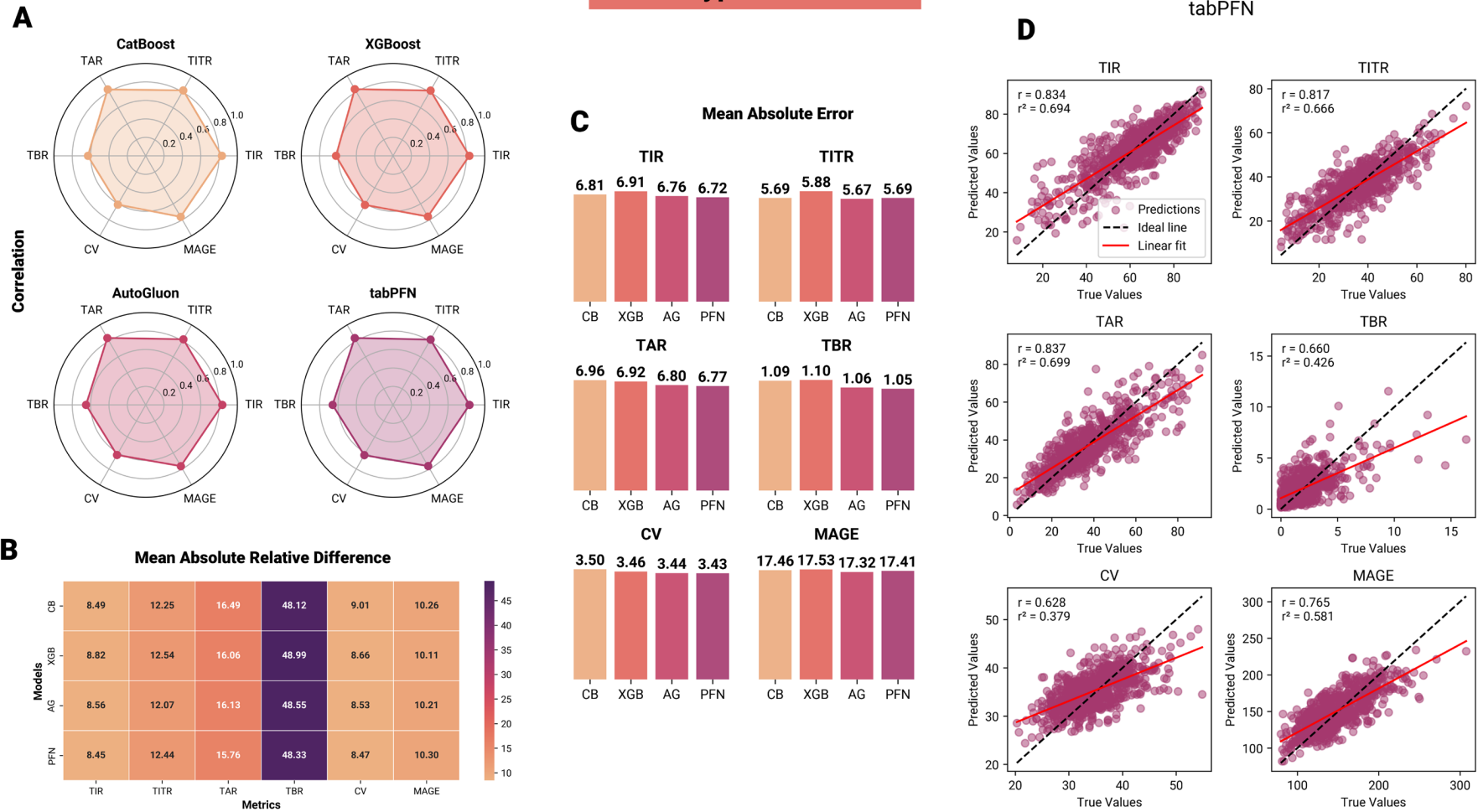

*Figure 2* – Performance of machine learning models in predicting glycemic control metrics in individuals with type 1 diabetes. (A) Radar plots of correlation coefficients between predicted and observed values across six metrics: time in range (TIR), time below range (TBR), time above range (TAR), time in tight range (TITR), coefficient of variation (CV), and mean amplitude of glycemic excursions (MAGE). (B) Heatmap of absolute relative difference (ARD) for each model–metric combination. (C) Bar plots of mean absolute error (MAE) across models and metrics. (D) Scatter plots of predicted versus observed values for tabPFN, including Pearson correlation coefficients (r) and coefficients of determination ($r^2$), with ideal (dashed) and linear fit (solid red) lines.

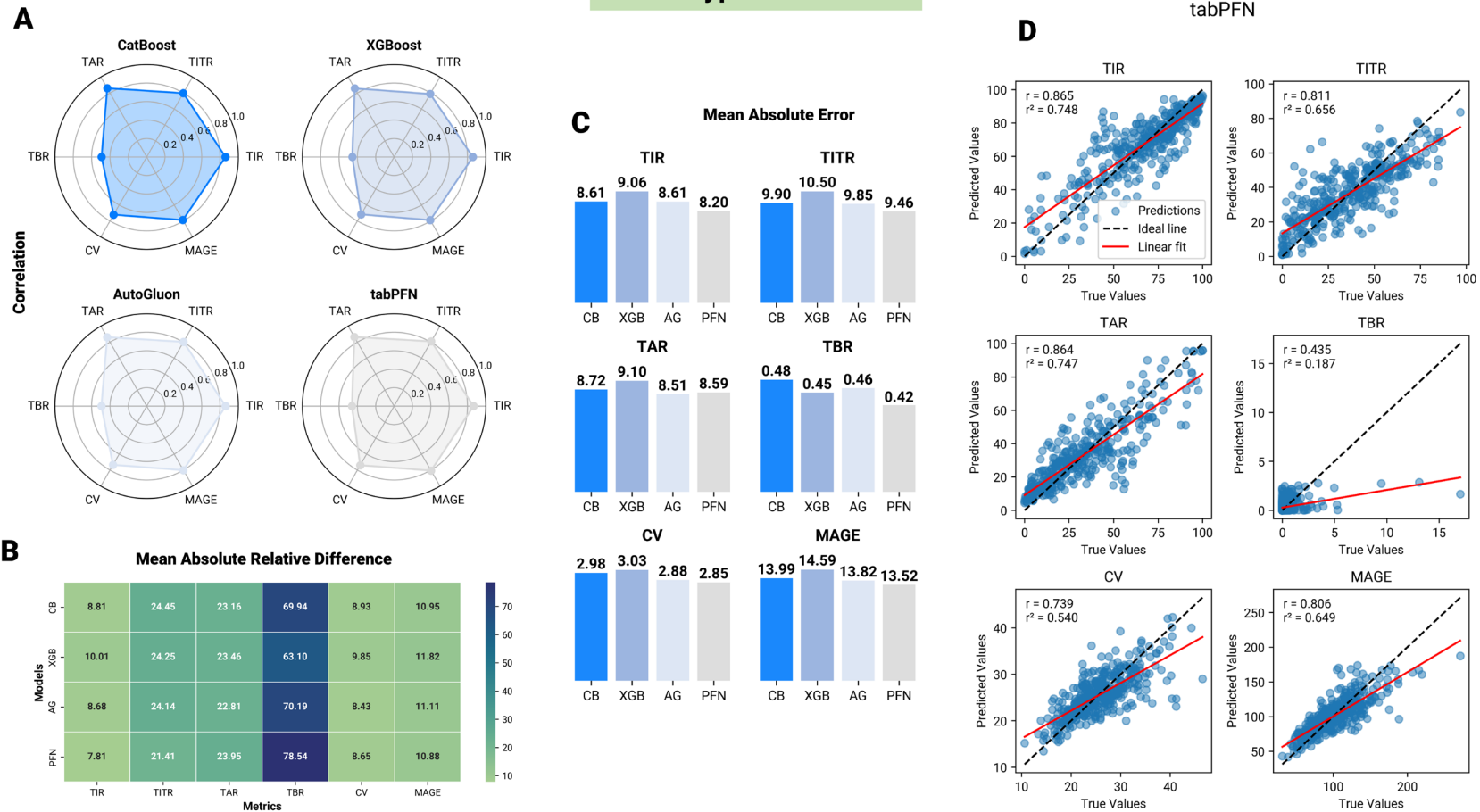

*Figure 3* – Performance of machine learning models in predicting glycemic control metrics in individuals with type 2 diabetes. (A) Radar plots of correlation coefficients between predicted and observed values across six metrics: time in range (TIR), time below range (TBR), time above range (TAR), time in tight range (TITR), coefficient of variation (CV), and mean amplitude of glycemic excursions (MAGE). (B) Heatmap of absolute relative difference (ARD) for each model–metric combination. (C) Bar plots of mean absolute error (MAE) across models and metrics. (D) Scatter plots of predicted versus observed values for tabPFN, including Pearson correlation coefficients (r) and coefficients of determination ($r^2$), with ideal (dashed) and linear fit (solid red) lines.

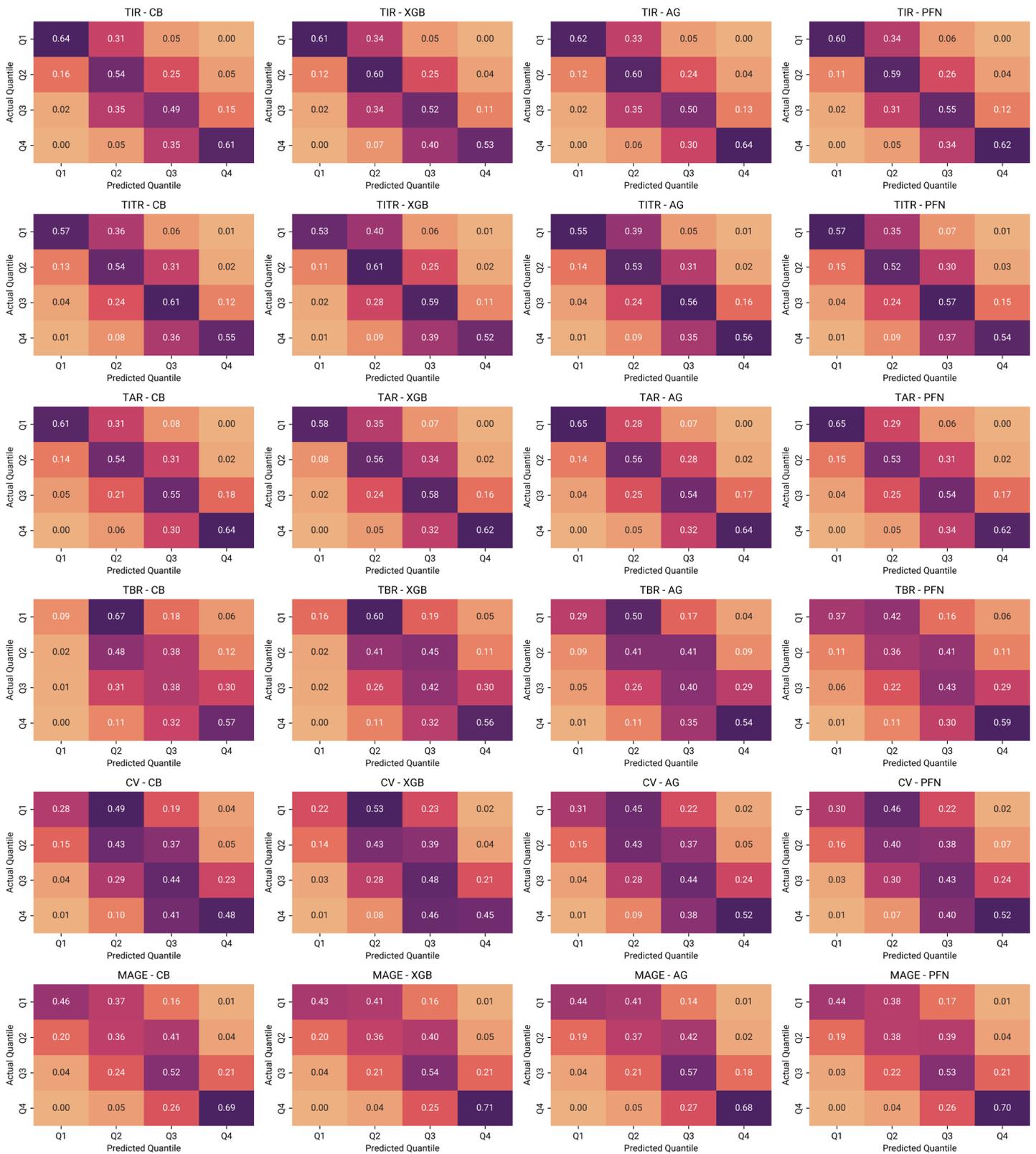

*Figure 4* – Confusion matrices of quantile predictions for glycemic control metrics in type 1 diabetes. Each panel displays the distribution of predicted versus actual quartiles (Q1–Q4) for one glycemic outcome (TIR, TITR, TAR, TBR, CV, MAGE) across four models (CatBoost[CB], XGBoost[XGB], AutoGluon[AG], tabPFN[PFN]). Values indicate the proportion of cases within each quantile, with darker shading reflecting higher accuracy along the diagonal.

# Supplementary Material

## LOCF Results
**Type 1 Diabetes**

|  | TIR | TITR | TAR | TBR | CV | MAGE |
|---|---|---|---|---|---|---|
| MSE | 382.154647 | 244.511850 | 399.370854 | 6.788063 | 47.254767 | 1731.177670 |
| R | 0.257668 | 0.251778 | 0.259886 | 0.138204 | 0.217457 | 0.257455 |
| R² | -0.482992 | -0.495208 | -0.478706 | -0.723549 | -0.563545 | -0.483959 |
| MAE | 14.329100 | 11.748304 | 14.568333 | 1.698437 | 5.238879 | 31.342716 |
| MRE_pct | 16.751640 | 24.797495 | 29.394608 | 70.796784 | 12.009752 | 16.590034 |

**Table S1.** Performance metrics for the LOCF prediction approach applied to weekly targets (TIR, TITR, TAR, TBR, CV, MAGE) in type 1 diabetes subjects. Metrics include mean squared error (MSE), correlation coefficient (r), coefficient of determination (R²), mean absolute error (MAE), and mean absolute relative difference (MRE% /MARD).

## LOCF Results
**Type 2 Diabetes**

|  | TIR | TITR | TAR | TBR | CV | MAGE |
|---|---|---|---|---|---|---|
| MSE | 704.261625 | 569.445602 | 714.271000 | 3.705424 | 54.004826 | 1569.563283 |
| R | 0.330777 | 0.381855 | 0.328288 | -0.004903 | 0.196983 | 0.218994 |
| R² | -0.339332 | -0.235916 | -0.344371 | -1.010139 | -0.604653 | -0.561446 |
| MAE | 18.660000 | 18.242821 | 18.852544 | 0.700151 | 5.494761 | 28.137160 |
| MRE_pct | 17.882140 | 42.133969 | 48.603352 | 100.000000 | 16.565077 | 21.504511 |

**Table S2.** Performance metrics for the LOCF prediction approach applied to weekly targets (TIR, TITR, TAR, TBR, CV, MAGE) in type 2 diabetes subjects. Metrics include mean squared error (MSE), correlation coefficient (r), coefficient of determination (R²), mean absolute error (MAE), and mean absolute relative difference (MRE% /MARD).

# Predicted versus observed values for LOCF
**Type 2 Diabetes**

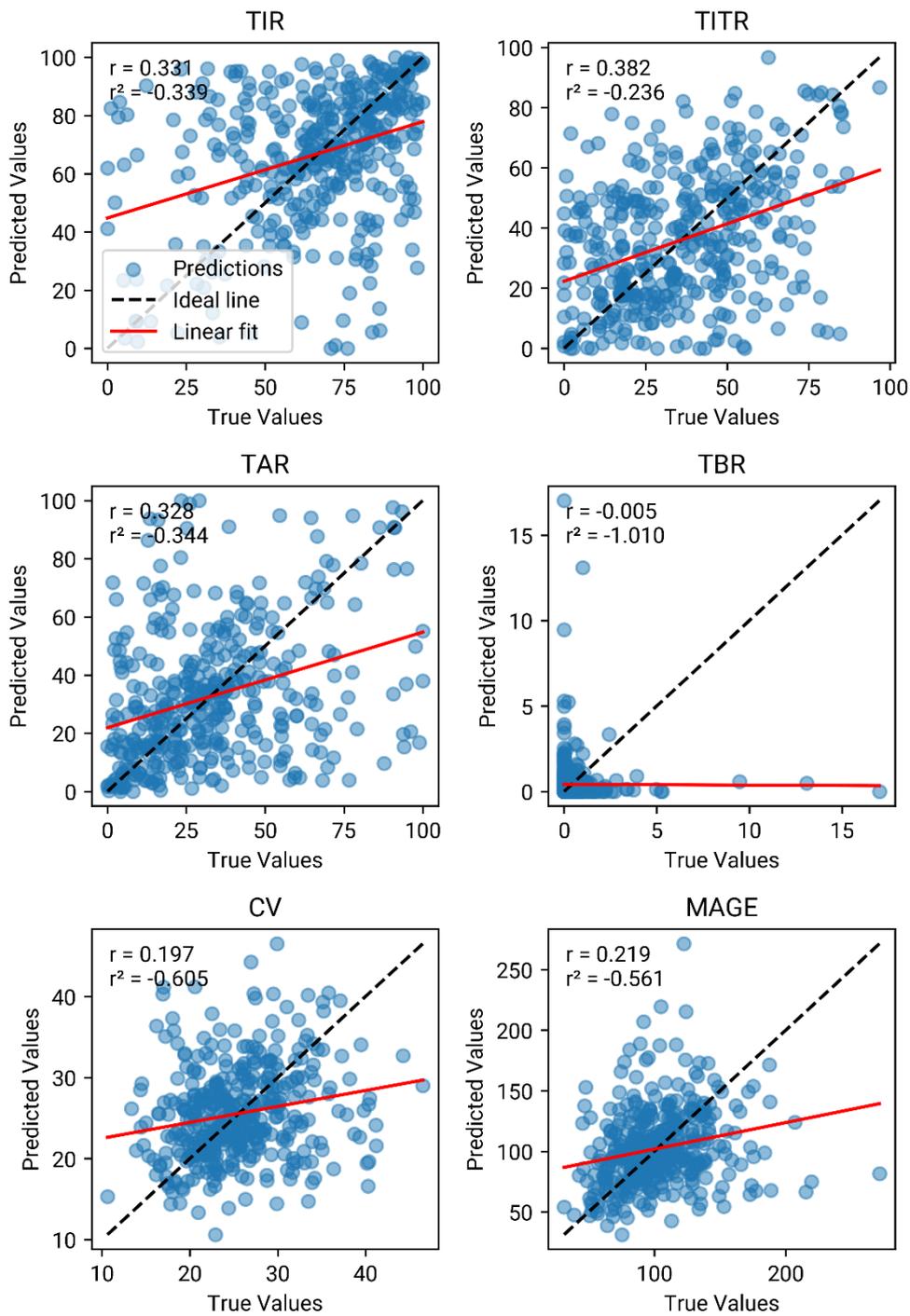

**Figure S1.** Scatter plots of predicted versus observed values for LOCF, including Pearson correlation coefficients (r) and coefficients of determination (r²), with ideal (dashed) and linear fit (solid red) lines.